\documentstyle[epsfig,psfig]{elsart}

\begin{document}
\begin{frontmatter}
\title{Relativity and Magnetism in Ni-Pd and Ni-Pt Alloys}
\author{Prabhakar P. Singh}
\address{Department of Physics \\
Indian Institute of Technology\\
Powai, Mumbai 400 076, India\\
}
\begin{abstract}
 We show that the differences in the magnetic properties of Ni-Pd and Ni-Pt 
alloys arise mainly due to relativity.  In particular, we find that the 
local magnetic moment of Ni {\it increases} with the addition of Pd in Ni-Pd 
while it {\it decreases} with the addition of Pt in Ni-Pt, 
as found experimentally,  
{\it only if relativity is present}.  Our analysis is 
based on the effects of relativity on (i) the spin-polarized densities 
of states of Ni, (ii) the splitting of  majority and minority spin 
$d$-band centers of Ni, and (iii) the separation between $s$-$d$ band 
centers of Pd and Pt in Ni-Pd and Ni-Pt alloys.
\end{abstract}

\end{frontmatter}
\section{Introduction}
\begin{figure}
\centering
\psfig{file=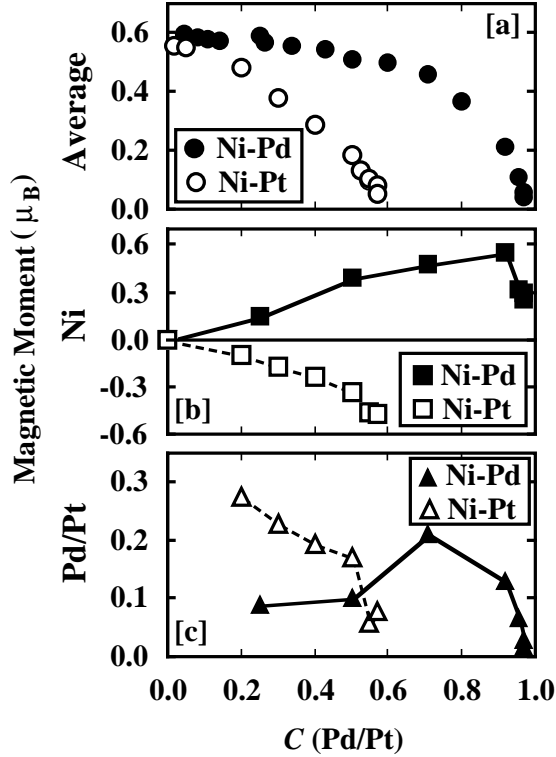, height=15cm}
\caption{The experimental (a) average magnetic moment, (b) local magnetic 
moment at Ni site, plotted with respect to the bulk Ni magnetic moment of 
0.616 $\mu_B$, and (c) local magnetic moment at Pd (Pt) site in Ni-Pd and 
Ni-Pt alloys}
\label{xmm}
\end{figure}

 The magnetic properties of alloys of $3d$ transition metals Fe, Co and Ni
with the nearly magnetic $4d$ Pd and $5d$ Pt show a wide range of behavior 
\cite{cad87,gub92,cab70,par79}.   
The variation in the magnetic properties of these alloys, as one goes from 
Fe-Pd(Pt) to Co-Pd(Pt) and then to Ni-Pd(Pt), can be attributed to the 
change in the number of valence electrons of one of the constituent atoms 
namely the $3d$ atoms.  However, the change in the magnetic properties of 
these alloys, as one replaces Pd by Pt, is not obvious because both Pd and 
Pt have the same number of valence electrons.  As an example, we 
show in Fig. \ref{xmm} the experimentally determined average magnetic moments 
and the local magnetic moments of Ni, Pd and Pt in Ni-Pd \cite{cad87,cab70} 
and Ni-Pt \cite{cad87,par79} alloys.  
From experiment \cite{cab70,par79} it is found that the addition of 
Pd to bulk Ni {\it increases} the magnetic moment of Ni 
(reaching a maximum at 
about 90\% Pd), whereas the addition of Pt to bulk Ni {\it decreases} the
magnetic moment of Ni.

 Earlier work \cite{par80,dah85} on the magnetic properties of Ni-Pd and 
Ni-Pt alloys used 
parametrized local environment models to describe the magnetism in Ni-Pd and 
Ni-Pt alloys. The local environment models incorporated the changes induced 
due to the chemical environment as well as the magnetic environment.  Recent 
work \cite{osw86,ste87,pis90,pis92}, based on the local spin density 
functional method, have 
simply calculated 
the magnetic properties of Ni-Pd and Ni-Pt alloys without trying to 
understand the electronic mechanism responsible for the differences 
in their magnetic properties.  The present study is, therefore, 
intended to improve our 
understanding of the reasons that lead to differences in the magnetic 
properties of Pd-based alloys and Pt-based alloys.  In particular, using 
Ni-Pd and Ni-Pt alloys as examples, we explain the reasons for the 
differences in their magnetic properties.

 The differences in the magnetic properties of Ni-Pd and Ni-Pt alloys 
are  dictated by the electronic structure of $4d$ Pd and $5d$ Pt atoms 
and their subsequent hybridization with Ni atoms.  
Since relativity is more important for heavier elements, the differences 
in the electronic  structure of Pd and Pt atoms are mainly due to relativity.  
Thus it is possible that the magnetism in Ni-Pt alloys is of relativistic 
origin which, in turn, may explain its anomalous behavior {\it vis-\'a-vis}
Ni-Pd alloys.

\section{Computational Details}
 In this paper we examine the effects of relativity, by including 
the so-called mass-velocity and Darwin terms, on the electronic 
structure of ordered Ni-Pd 
({\it L}1$_2$ Ni$_3$Pd, {\it L}1$_0$ NiPd and {\it L}1$_2$ NiPd$_3$) 
and Ni-Pt 
({\it L}1$_2$ Ni$_3$Pt, {\it L}1$_0$ NiPt and {\it L}1$_2$ NiPt$_3$) 
alloys.   
The electronic structure of Ni-Pd and Ni-Pt alloys are obtained by 
carrying out spin-polarized, charge self-consistent 
calculations using the linear muffin-tin orbital (LMTO) method in the 
atomic-sphere approximation (ASA) \cite{and75,and84,skr84}, including the 
combined correction terms \cite{and75}.  
The calculations are carried out with the non relativistic Schr\"odinger 
equation as well as with the scalar-relativistic Dirac equation. 
In our calculations, the ratio of the Ni-atomic sphere radii, $R_{Ni}$, and 
the average Wigner-Seitz (WS) radii $R_{WS}$, in Ni-Pd and Ni-Pt 
alloys are chosen to make the respective atomic spheres charge neutral. We 
find that the  charge neutrality is obtained for 
$R_{Ni}/R_{WS}$ = 0.98 (0.97), 0.97 (0.95)  and 0.96 (0.93) 
for Ni$_3$Pd (Ni$_3$Pt),
 NiPd (NiPt) and NiPd$_3$ (NiPt$_3$) alloys, respectively.  
The {\bf k}-space 
integrations are carried out with sufficient number of {\bf k} points in 
the irreducible wedge of the corresponding Brillouin zone to ensure 
the convergence of the magnetic moment \cite{sin98}.  The results, 
described below, correspond to the calculated equilibrium volume in 
each case.
\begin{figure}
\centering
\psfig{file=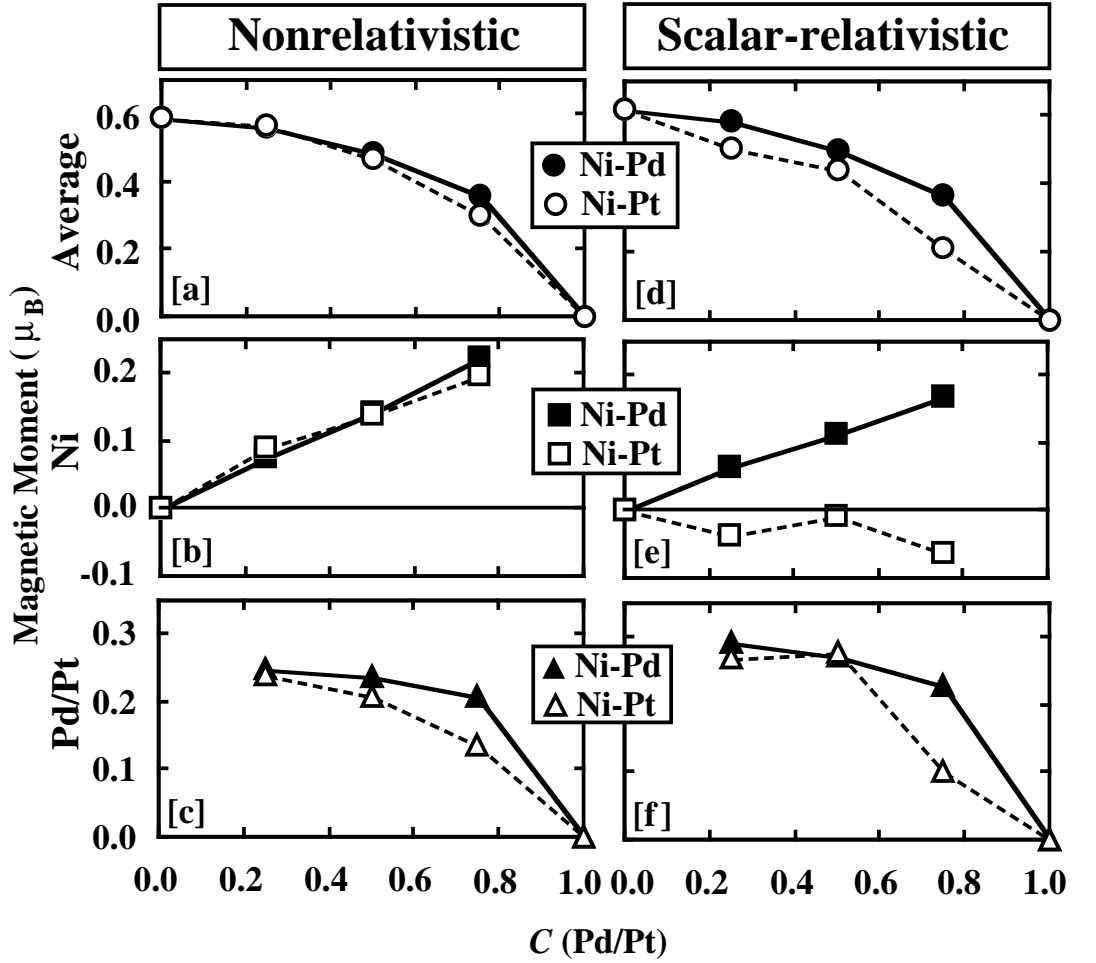,height=15cm,angle=-90}
\caption{The non relativistically and scalar-relativistically calculated 
(a, d) average magnetic moment, (b, e) local magnetic moment at Ni site, 
plotted with respect to the calculated bulk Ni magnetic moment in each case, 
and 
(c, f) local magnetic moment at Pd (Pt) site in Ni-Pd and Ni-Pt alloys.}
\label{tmm}
\end{figure}

\section{Results and Discussion}
 The most significant of our results are shown in Fig. \ref{tmm}, 
where we show the 
average as well as the local magnetic moments at the Ni and Pd (Pt) sites of 
Ni-Pd (Pt) alloys calculated non relativistically, Fig. \ref{tmm}(a)-(c), and 
scalar-relativistically, Fig. \ref{tmm}(d)-(f), as a function of Pd (Pt) 
concentration.  
The average magnetic moment calculated non relativistically, Fig. \ref{tmm}(a), 
for both Ni-Pd and Ni-Pt alloys are very similar, 
and they decrease as a function of Pd or Pt concentration, respectively.  
The magnetic moment at 
the Ni site, as shown in Fig. \ref{tmm}(b), increases substantially with 
increase 
in Pd or Pt concentration. The magnetic moment 
at the Pd or Pt site, Fig. \ref{tmm}(c), does not vary by much and is 
relatively small. 
Thus Figs. \ref{tmm}(a)-(c) clearly show that if we use the non relativistic 
Hamiltonian to describe the electronic 
structure then both Ni-Pd and Ni-Pt alloys display similar magnetic 
behavior.  However, as shown in Figs. \ref{tmm}(d)-(f), the calculated 
average as 
well as local magnetic moments are quite different for Ni-Pd 
and Ni-Pt alloys if 
we use the scalar-relativistic Hamiltonian to describe their electronic 
structure.  In particular, we find from Fig. \ref{tmm}(e) that the magnetic 
moment at the Ni site in Ni-Pd 
alloys increases with Pd concentration, whereas it decreases with 
increasing Pt concentration in Ni-Pt alloys.  As a function of concentration 
the change in the local magnetic moment at the Pd or Pt site, shown in 
Fig. \ref{tmm}(f), is small.  {\it Thus a comparison of our non relativistic 
and scalar-
relativistic 
results clearly shows that the differences in the magnetic properties of 
Ni-Pd and Ni-Pt alloys are, to a large extent, determined by relativity.}  
We next compare our calculated magnetic moments of Ni-Pd and Ni-Pt alloys 
with the calculations 
of others and the experimental values.

The calculated magnetic moment of fcc Ni, which changes from 0.59 $\mu_B$ 
to 0.62 $\mu_B$ with the inclusion of relativistic terms, compares very well 
with the experimental value of 0.616 $\mu_B$ and the non relativistic 
calculation of Ref. \cite{mor78}.  Note that our calculation 
ignores the orbital contribution to the magnetic moment of Ni which is 
expected to be around 0.05 $\mu_B$.  The average magnetic moments, calculated 
scalar-relativistically, for Ni$_3$Pd, NiPd and NiPd$_3$  are 0.58 $\mu_B$, 
0.50 $\mu_B$ and 0.37 $\mu_B$ respectively.  Since we find that the 
substitutional disorder does not change the magnetic moment of Ni-Pd alloys 
at 25\%, 50\% and 75\% concentrations of Pd by much \cite{sin98u}, 
our calculated values 
can be compared with the experimental values \cite{cab70},
 0.59 $\mu_B$, 0.51 $\mu_B$ 
and 0.46 $\mu_B$, obtained for disordered 
Ni-Pd alloys at 25\%, 50\% and 71\% concentrations of Pd respectively. 
The calculated local moments at the Ni (Pd) site in Ni-Pd alloys is somewhat 
smaller (larger) than 
the experimentally determined values, which can be due to uncertainty 
involved in site-decomposing the average magnetic moment.  For Ni$_3$Pt 
the scalar-relativistic calculations show the average magnetic moment 
to be 0.50 $\mu_B$ with the local magnetic moments at the Ni and Pt sites 
being 0.58 $\mu_B$ and 0.27 $\mu_B$ respectively.  The corresponding 
experimental 
values \cite{par79,par80,alb79} 
are 0.43 $\mu_B$ (average), 0.49 $\mu_B$ (Ni) and 0.25 $\mu_B$ (Pt) 
respectively.  The results of Refs.\cite{pis90,pis92} on Ni$_3$Pt are 
based on self-
consistent local spin density approximation using the LMTO method but 
the value of the local magnetic moment of Ni given in the two references 
are different.

 To further understand the electronic mechanism responsible for the 
differences in the magnetic properties of Ni-Pd and Ni-Pt alloys, we 
examine (i) the spin-polarized densities of states (DOS) of Ni, 
(ii) the separation between 
majority and minority spin $d$-band centers \cite{skr84} of Ni, 
$\Delta C^{Ni}_{d\uparrow - d\downarrow}$,
and (iii) the separation between $s$- and $d$- band centers of Pd and Pt, 
$\Delta C^{Pd/Pt}_{s\uparrow - d\uparrow}$ 
in Ni-Pd and Ni-Pt alloys.
\begin{figure}
\centering
\psfig{file=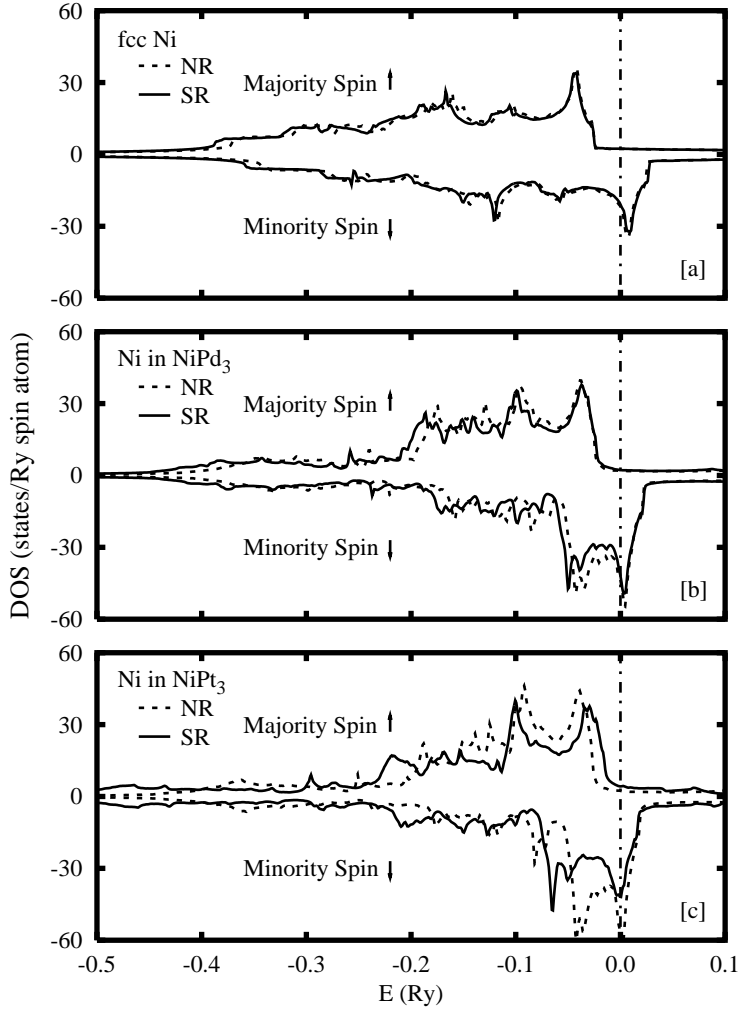,height=15cm}
\caption{The spin-polarized densities of states of Ni, calculated 
non relativistically (NR) and scalar-relativistically (SR), in (a) fcc Ni, 
(b) {\it L}1$_2$ NiPd$_3$, and (c) {\it L}1$_2$ NiPt$_3$.}
\label{dos}
\end{figure}

\subsection{Spin-Polarised Densities of States}
In Fig. \ref{dos} we show the spin-polarized DOS at the Ni site in fcc Ni,
{\it L}1$_2$ NiPd$_3$ and {\it L}1$_2$ NiPt$_3$ calculated with 
the non relativistic 
and the scalar-relativistic Hamiltonian.  As expected, the relativistic 
effects on the DOS of the elemental Ni, shown in Fig. \ref{dos}(a), are 
very small.  
Since relativity is more important for 
Pd than for Ni, its effect on the DOS at the Ni site in NiPd$_3$ is more 
pronounced than in fcc Ni as shown in Fig. \ref{dos}(b).  We find that 
in NiPd$_3$ the inclusion of 
relativity leads to a decrease in the magnetic moment at the Ni site by 
0.03 $\mu_B$.  However, in the case of NiPt$_3$, as shown in 
Fig. \ref{dos}(c), the effect
on the magnetic moment is an order of magnitude larger than for NiPd$_3$.  
The relativity reduces the magnetic moment at the Ni site in NiPt$_3$ 
by 0.24 $\mu_B$, 
i.e., from 0.79 $\mu_B$ to 0.55 $\mu_B$.  Once again the change in the 
magnetic moment at the Pt site is negligible in comparison.  

\begin{figure}
\centering
\psfig{file=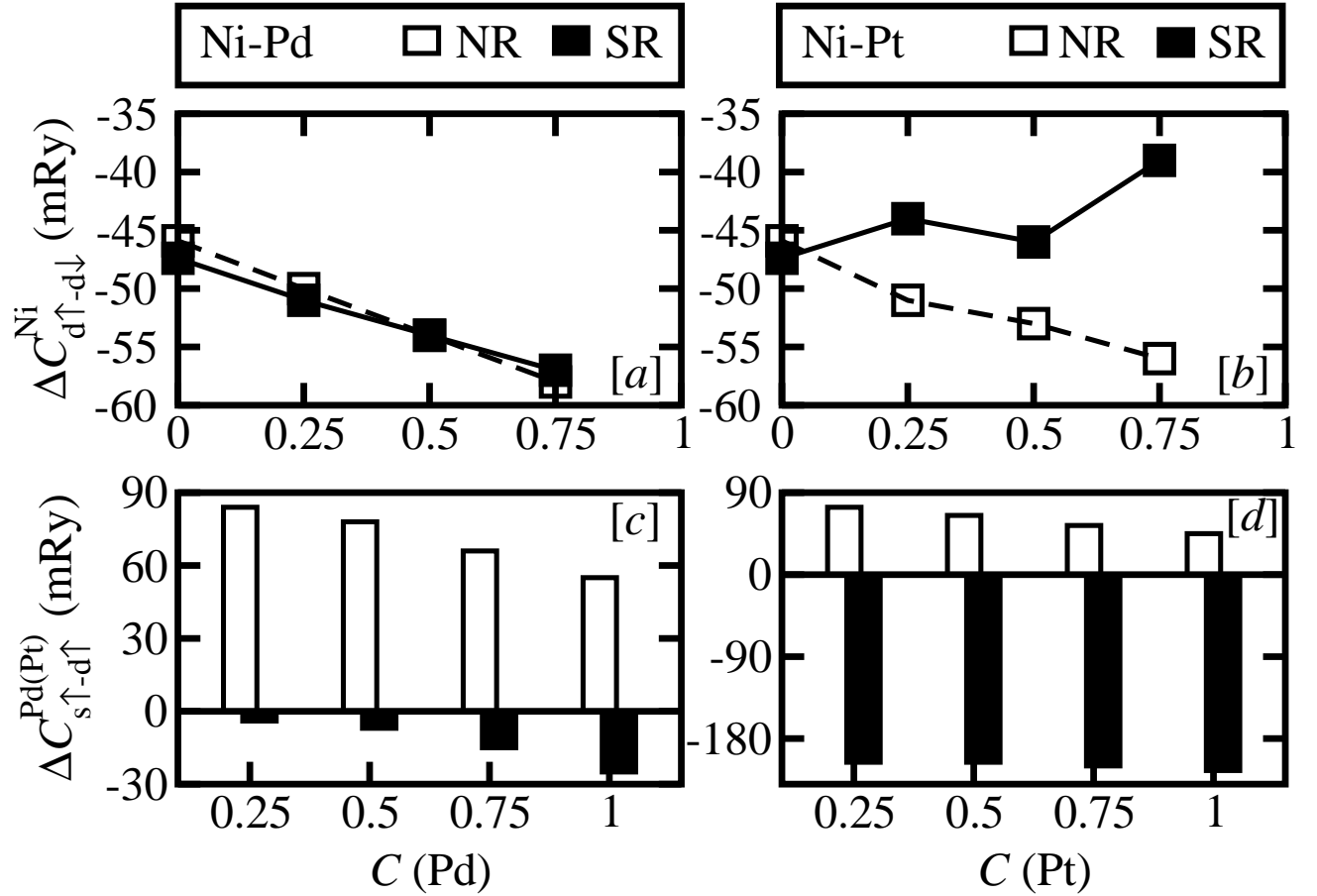,height=15cm,angle=-90}
\caption{The non relativistic (NR) and scalar-relativistic (SR) exchange-
induced splitting of majority spin and minority spin $d$-band centers of Ni in 
(a) Ni-Pd and (b) Ni-Pt alloys.  Figures (c) and (d) show the separation 
between majority spin
$s$ and $d$ band centers of (c) Pd in  Ni-Pd and (d) Pt in 
Ni-Pt alloys.}
\label{split}
\end{figure}

\subsection{Separation Between Majority and Minority Spin $d$-band Centers}
 A more quantitative explanation for the changes in the magnetic moments 
due to relativity can be obtained by examining 
the separation between 
majority spin and minority spin $d$-band centers of Ni   
($\Delta C^{Ni}_{d\uparrow - d\downarrow}$) in Ni-Pd, shown in 
Fig. \ref{split}(a), 
and Ni-Pt, shown in 
Fig. \ref{split}(b), alloys.  As can be seen from Fig. 4(a), the 
exchange-induced splitting
of the $d$-band increases  with the addition of Pd in Ni-Pd alloys for 
calculations done with or without relativity.  The increased splitting leads 
to an increase in the local magnetic moment at the Ni site, as observed 
experimentally.  It is interesting to note that the inclusion of relativity 
produces no net change in the exchange-induced splitting at the Ni site in 
equiatomic NiPd.  On the other hand, in Ni-Pt alloys we find that 
relativity substantially reduces the exchange-induced splitting at the Ni 
site leading to a decrease in the local magnetic moment of Ni. For example, 
in Ni$_3$Pt the separation between $d$-band centers reduces from 
51 mRy to 44 mRy and the corresponding reduction in the local magnetic moment 
is from 0.68 $\mu_B$ to 0.58 $\mu_B$, in agreement with experiment.  With 
increasing Pt concentration the relativistic effects become more dominant  
which further reduces the splitting, as is the case for NiPt$_3$.

\subsection{Separation Between $s$ and $d$ Band Centers}
 It is clear that the differences in the magnetic properties of Ni-Pd and Ni-Pt alloys are brought about by relativity through its effect on Pd and Pt atoms.  
To see how relativity affects the electronic structure of Pd and Pt atoms, 
we show in Figs. \ref{split}(c)-(d) the separation between the majority spin 
$s$- and $d$-band centers of 
Pd and Pt atoms as a function of concentration.  
We know that the most dominant effect of 
relativity is to lower the $s$ potential.  The lowering of $s$ potential 
causes (i) the $s$-wavefunction to contract leading to a contraction of the 
lattice \cite{sin93a,sin96,sin94}, and (ii) increased $s$-$d$ 
hybridization which results in 
electron transfer from $d$ to $s$ \cite{sin94}.  We see from 
Figs. \ref{split}(c) 
and \ref{split}(d) that the 
change in $s$-$d$ separation is almost an order of magnitude more in Pt 
than in Pd.  For example, the $s$-$d$ separation for Pd in NiPd$_3$ changes 
from +66 mRy to -15 mRy, whereas for Pt in Ni$_3$Pt it changes from +74 mRy 
to -206 mRy.  
Thus the contraction of the $s$ wavefunction of Pt and 
the subsequent $s$-$d$ hybridization  must be responsible for 
reducing  the local 
magnetic moment at the Ni site.  

\begin{figure}
\centering
\psfig{file=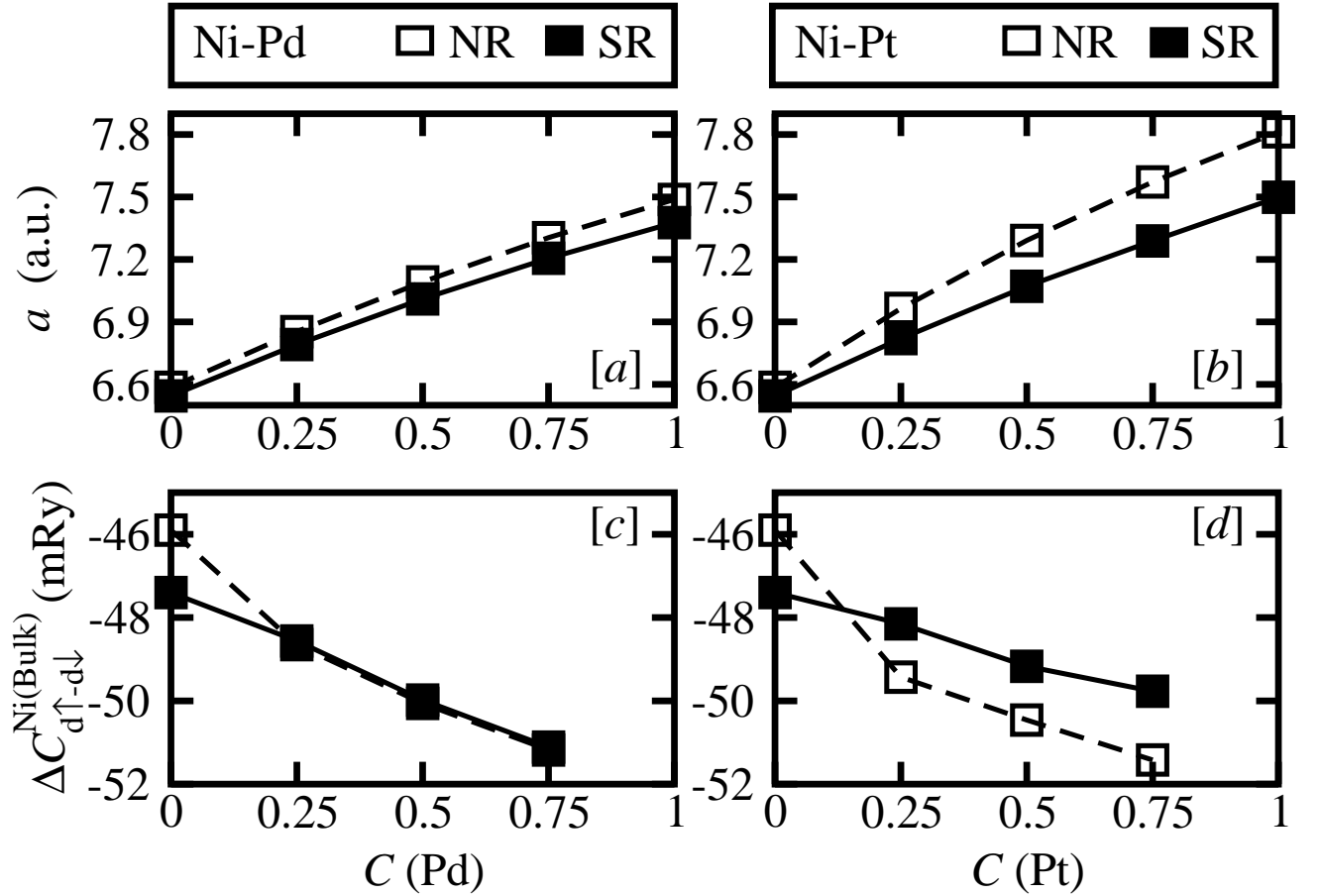,height=15cm,angle=-90}
\caption{The non relativistically (NR) and scalar-relativistically (SR) 
 calculated equilibrium lattice constants for 
(a) Ni-Pd and (b) Ni-Pt alloys.  Figures (c) and (d) show the 
exchange-induced splitting between the  
majority  and minority  
$d$ band centers in fcc Ni, calculated  using  the average Wigner-Seitz radii 
in bulk Ni equal to  the radii of the 
charge neutral Ni atomic spheres in  (c) Ni-Pd and (d)  
Ni-Pt alloys.}
\label{lat}
\end{figure}

 The effects of the contraction of $s$ wavefunction in Ni-Pd and Ni-Pt alloys can be 
clearly seen in Figs. \ref{lat}(a)-(b), where we show the equilibrium lattice constants calculated 
non relativistically and scalar-relativistically for these alloys. We
find that the change in the lattice constant due to relativity is much more 
for Ni-Pt alloys than for Ni-Pd alloys.  For example, in NiPt$_3$ relativity 
reduces the lattice constant by around 0.3 a.u., whereas the corresponding change in NiPd$_3$ is only about 0.1 a.u.. Such a drastic reduction in the 
lattice constant due to relativity effectively put the Ni sublattice under 
strain in Ni-Pt 
alloys  which, in turn, reduces the magnetic moment at the Ni site. To 
see how relativity-induced  strain in Ni sublattice leads to a reduction in the 
local magnetic moment in 
Ni-Pt alloys, we have used the radii of the charge-neutral Ni atomic spheres 
(R${Ni}$) in these alloys as the average Wigner-Seitz radii (R$_{WS}$) to calculate, self-consistently, the electronic structure of 
fcc Ni.  The corresponding results, in terms of the  separation between 
majority and 
minority $d$-band centers  in fcc Ni, 
($\Delta C^{Ni(Bulk)}_{d\uparrow - d\downarrow}$), are shown in Figs. 
\ref{lat}(c)-(d). As shown in Fig. \ref{lat}(c), the relativistic terms have  
a very small  effect  on the 
exchange splitting in Ni-Pd alloys. However, 
in Ni-Pt alloys the exchange splitting is {\it reduced} with increasing Pt 
concentration, as can be seen from Fig. \ref{lat}(d). The reduced exchange 
splitting leads to diminished local magnetic moment at the Ni site in 
Ni-Pt alloys, consistent with Fig. \ref{split}(b) and the experiments.

 We like to point out that  the present study can be further improved  by 
including (i) the full-potential instead of spherically symmetric potential 
used in the ASA, and (ii) the spin-orbit terms.  
Also, as Ni and Pd form a 
ferromagnetic fcc solid solution throughout the concentration range 
while Ni and Pt form a ferromagnetic fcc solid solution for Pt concentration 
below 0.6, an approach based on the study of disordered alloys can 
lead to a more accurate description of the magnetic properties of 
these alloys.  Thus one can use the Korringa-Kohn-Rostoker coherent-potential 
approximation in the ASA (KKR-ASA CPA) \cite{sin93a,sin93b} to describe the 
electronic structure of disordered Ni-Pd and Ni-Pt alloys.  However, 
for a reliable description of the magnetic properties of 
disordered Ni-Pd and Ni-Pt alloys, 
the overlap errors associated with the ASA must be corrected \cite{sin98} 
which is not 
possible in the present implementation of the KKR-ASA CPA 
method \cite{sin93a}.  We emphasize that the improvements outlined above are 
unlikely to change the main results of the present study because of the 
robustness of the relativistic effects. 

\section{Conclusions}
 In conclusion, we have shown that the differences in the magnetic 
properties of Ni-Pd and Ni-Pt alloys arise due to relativity.  In 
particular, relativity ensures that  the local magnetic moment of Ni 
increases with addition of Pd in Ni-Pd while it decreases with addition of 
Pt in Ni-Pt, consistent with experiment.  We also find that the decrease in 
the local magnetic moment of Ni in Ni-Pt alloys is facilitated by relativity 
through lowering of the $s$ potential of Pt, which leads to a  contraction of 
the $s$ wavefunction and an increase in $s$-$d$ hybridization.

\end{document}